\documentclass[aps,prb,twocolumn,superscriptaddress,notitlepage]{revtex4-1}
\usepackage{graphicx}
\usepackage[caption=false]{subfig}
\usepackage{float}
\usepackage{natbib}
\usepackage{xcolor}
\usepackage{xr}
\usepackage{amsmath}
\usepackage{amssymb}
\usepackage{array}
\usepackage{wasysym}
\usepackage{color,soul}
\usepackage{braket}
\usepackage{verbatim}
\usepackage{hyperref}
\usepackage{gensymb}
\usepackage{url}
\usepackage{tabularx}
\usepackage{filecontents}
\usepackage{textgreek}
\usepackage{gensymb}
\newcommand{\LMP}{LiMoOP$_2$O$_7$}
\newcommand{\AMP}{\textit{A}MoOP$_2$O$_7$}

\begin{document}

\title{Magnetic properties and excitation spectrum of the $S = 1/2$ three-legged spin ladder, LiMoOP$_2$O$_7$}

\author{Aly~H.~Abdeldaim}
\email{aly.abdeldaim@diamond.ac.uk}
\affiliation{Diamond Light Source, Harwell Science and Innovation Campus, Didcot, OX11 0DE, United Kingdom}
\affiliation{School of Chemistry, University of Birmingham, Edgbaston, Birmingham B15 2TT, United Kingdom}
\affiliation{ISIS Neutron and Muon Source, Science and Technology Facilities Council, Didcot OX11 0QX, United Kingdom}
\author{Alexander~A.~Tsirlin}
\affiliation{Felix Bloch Institute for Solid-State Physics, Leipzig University, 04103 Leipzig, Germany}
\author{Jacques~Ollivier}
\affiliation{Institut Laue-Langevin, 71 Avenue des Martyrs, CS20156, 38042 Grenoble Cédex 9, France}
\author{Clemens~Ritter}
\affiliation{Institut Laue-Langevin, 71 Avenue des Martyrs, CS20156, 38042 Grenoble Cédex 9, France}
\author{Robin~S.~Perry}
\affiliation{London Centre for Nanotechnology and Department of Physics and Astronomy,University College London, London WC1E 6BT, United Kingdom}
\affiliation{ISIS Neutron and Muon Source, Science and Technology Facilities Council, Didcot OX11 0QX, United Kingdom}
\author{Lucy~Clark}
\affiliation{School of Chemistry, University of Birmingham, Edgbaston, Birmingham B15 2TT, United Kingdom}
\author{Gøran~J.~Nilsen}
\email{goran.nilsen@stfc.ac.uk}
\affiliation{ISIS Neutron and Muon Source, Science and Technology Facilities Council, Didcot OX11 0QX, United Kingdom}
\affiliation{Department of Mathematics and Physics, University of Stavanger, 4036 Stavanger, Norway}
\date{\today}

\date{\today}

\begin{abstract}
\noindent
We report a comprehensive experimental study including density functional theory calculations detailing the magnetic ground state of the $S = 1/2$ Mo\textsuperscript{5+} pyrophosphate, \LMP, and its magnetic excitations. Neutron powder diffraction measurements confirm the $P2_1/n$ structural model which is comprised of a three-legged ladder of distorted Mo\textsuperscript{5+} octahedra. Magnetometry and specific heat reveal a long-range magnetic ordering transition at $T_\textrm{N} = 3.5$~K and a field-induced transition at $\mu_0 H_\textrm{c} = 1$~T. This zero-field long-range magnetic order was determined to be a commensurate $\kappa = 0$ order with an ordered moment that is reduced from the full $S = 1/2$ magnetic moment by $30\%$. Through inelastic neutron scattering measurements and density functional theory calculations, we define the minimal microscopic model and resolve the spin Hamiltonian up to the seventh leading exchange parameter, resulting in a model with dominant antiferromagnetic leg interactions, weaker ferromagnetic rung interactions, and a significant interlayer ferromagnetic coupling.

\end{abstract}
\maketitle
\section{Introduction}
\noindent The spatial dimensionality of the magnetic sublattice in a material plays a central role in determining its collective electronic behavior. This is evident when comparing, for example, the macroscopic singlet ground state realized in the $S = 1/2$ Heisenberg one-dimensional chain antiferromagnet\cite{Bethe1931,Hulthen1938,Mourigal2013} with the N\'eel order observed for its two-dimensional square lattice counterpart\cite{shannon2004finite}. Spin models in which the effective dimensionality can be tuned between these limits are therefore of particular interest for exploring its influence on the magnetic ground state.  

The $S = 1/2$ Heisenberg spin ladder model provides a natural framework for studying this dimensional crossover. It describes systems comprised of a finite number of one-dimensional spin chains that are coupled by transverse rung interactions. Given the large parameter space, the Heisenberg spin ladder model hosts a rich phase diagram that is primarily defined by the number of spin chains, or ladder legs, $n$, the sign of the exchange interactions, and the relative strength of the leg to rung interactions\cite{Vasiliev2018}. For antiferromagnetic (AFM) interactions and an even $n$, for example, the ground state is described as a collective singlet state with exponentially decaying short range correlations and a finite gap to the lowest lying triplet excited states. Odd $n$ AFM ladders, on the other hand, have interactions that are described by power-law decaying spin correlations and a gapless excitation spectrum. These ground states have motivated extensive theoretical and experimental investigations\cite{Dagotto1988,Storng1992,Hida1991,Watanabe1993,Shelton1996,Frischmuth1997,Imai1998,Azuma1994,Eccleston1998}, and the broader phase diagram continues to provide a setting for exploring a broad range of quantum phenomena including superconductivity\cite{Takahashi2015,Dagotto1992}, Tomonaga-Luttinger-liquid physics\cite{Sakai2022,Mingee2013,Thielemann2009}, and Bose-Einstein condensation\cite{Kono2017,Garlea2007}, amongst others.

\begin{figure*}
\centering
\includegraphics[width = \textwidth,keepaspectratio]{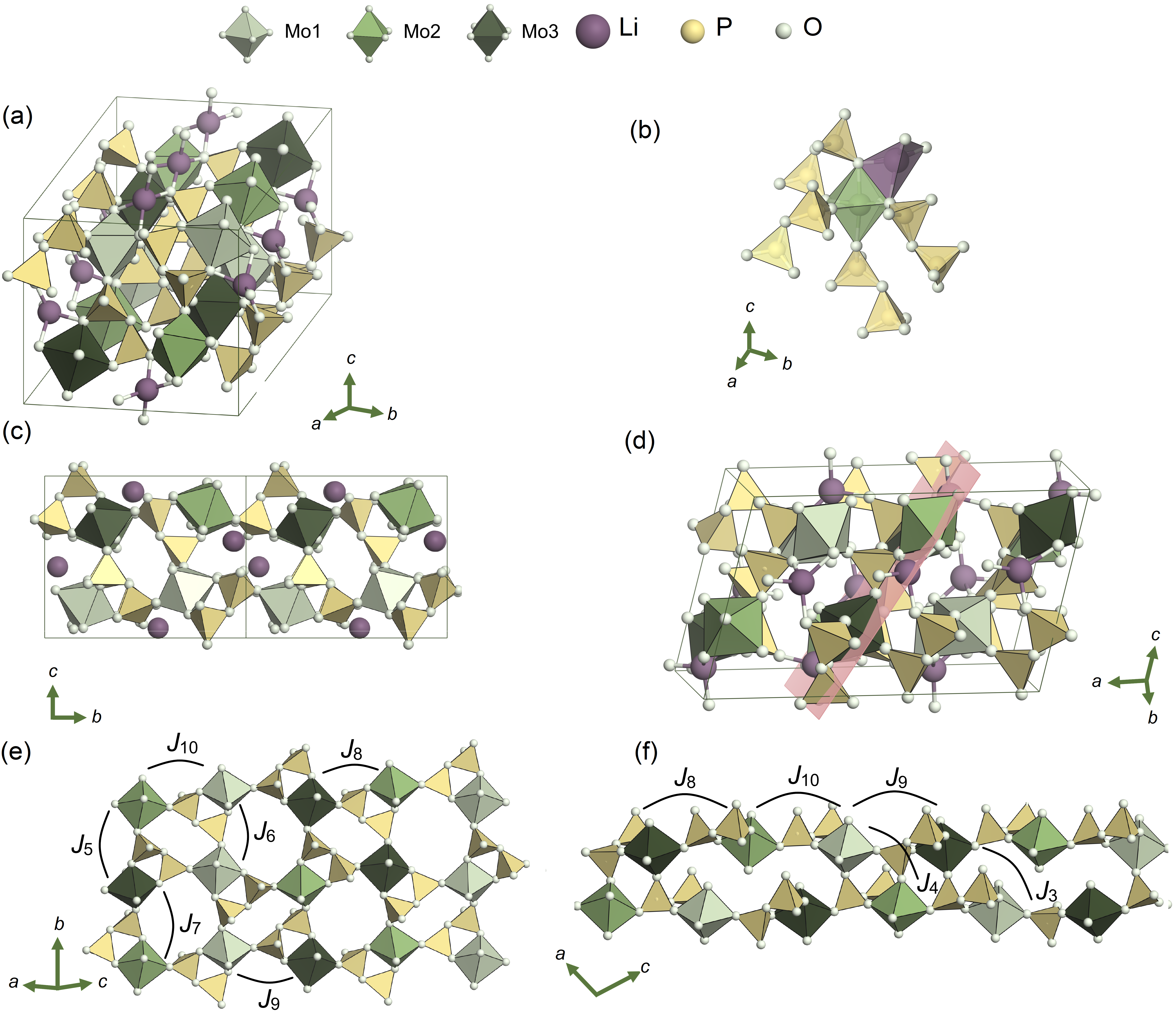}\caption{(a) Crystal structure of \LMP~viewed along the [111] direction. (b) All Mo positions coordinate four P$_2$O$_7$ groups and share edges with a LiO$_4$ tetrahedron leaving one free apex that (c) points towards an empty tunnel-like structure that runs along the $a$-axis. Another type of tunnel forms in which the Li cations run along the $a$-axis. The magnetic sublattice of \LMP~is comprised of (d-e) three-legged ladders that propagate along the [301] plane and are (f) coupled to the rest of the framework within the $ac$ plane. Figures generated using the \texttt{VESTA} visualization package\cite{vesta}.}
\label{figure1}
\end{figure*}

Within this context, we have examined the $S = 1/2$ Mo\textsuperscript{5+}-based pyrophosphates \textit{A}\textsuperscript{+}MoOP$_2$O$_7$ where \textit{A}\textsuperscript{+}$ = $ Li\textsuperscript{+}\cite{LMP}, Na\textsuperscript{+}\cite{NMP}, K\textsuperscript{+}\cite{KMP}, and Cs\textsuperscript{+}\cite{CMP}. Our previous work has established KMoOP$_2$O$_7$ as a near-ideal realization of the $S = 1/2$ Heisenberg one-dimensional chain antiferromagnet model\cite{Abdeldaim2023}. \LMP, in contrast, adopts a distinct crystal structure which forms a distorted three-legged spin ladder (Fig. \ref{figure1}) that we explore here through a combination of powder neutron diffraction and inelastic neutron scattering, magnetometry, and band structure calculations. We begin in Sec. \ref{sec:methods} by detailing our synthetic, experimental, and computational methods. In Sec. \ref{sec:results}, we confirm the crystal structure of \LMP, find commensurate long-range magnetic order at $T_\textrm{N} = 3.5$~K, and quantify the exchange parameters of the spin Hamiltonian. We finally discuss the exchange pathways in Sec. \ref{sec:discussion}, before concluding in Sec. \ref{sec:conclusions}.
%%%%%%%%%%%%%%%%%%%%%%%%%%%%%%%%%%%%%
%%%%%%%%%%%%%%%%%%%%%%%%%%%%%%%%%%%%%
\vspace{-2mm}
\section{Methods}
\label{sec:methods}
\noindent Polycrystalline samples of \LMP~were prepared following a modified version of a previously published solid-state method \cite{LMP}. In a typical reaction, MoO$_3$ (Alfa Aesar, 99.998\%), Mo (Alfa Aesar, powder, $<$150~$\mu$m, 99\%), (NH$_4$)$_2$HPO$_4$ (Alfa Aesar, 98\%+), and \textsuperscript{7}Li$_2$CO$_3$ (Sigma-Aldrich, $>$99\%) were combined in a molar ratio of 1.68:0.32:4:1 and wet-ground in isopropyl alcohol for 30~minutes in a Retsch planetary ball mill. The sample was then heated at 623~K for 12~hours in an alumina crucible before cooling to room temperature. The resulting dark-blue powder was manually ground and pressed into a pellet prior to sintering at 833~K for 24~hours in a tube furnace under flowing argon. Finally, the sample was washed with acetone, ground, repelletized, and heated at 833~K for 12~hours in an evacuated quartz ampoule backfilled with 300~mbar of argon gas to result in the bright blue final \LMP~product. All heating and cooling stages were done at rates of $Q_\textrm{H} =Q_\textrm{C} = 3$~K~min$^{-1}$.

Neutron powder diffraction data were collected for a 5~g sample of $^7$Li-enriched \LMP~at room temperature on the D2B diffractometer and at 10~K and 1.7~K on the D20 diffractometer at the Institut Laue-Langevin (ILL). Rietveld analysis of the crystal structure was performed using the \texttt{GSAS-II} software package\cite{gsasii} and magnetic structure refinements were conducted using \texttt{FULLPROF}\cite{fullprof} and \texttt{Mag2Pol}\cite{Qureshi2019}. A covalent form factor of Mo\textsuperscript{5+} was used and was calculated using density functional theory\cite{Tsirlinff}.

The temperature dependence of the magnetic susceptibility was measured in DC mode for a 17.47~mg sample between 1.8~K and 300~K at 0.1~T and 4~T using both field-cooled and zero-field cooled protocols on a Quantum Design MPMS3 SQUID magnetometer. The magnetization isotherm was measured between $-7$~T and $7$~T at 1.8~K. Temperature-dependent specific heat data were measured between 1.8~K and 300~K for a 2.69~mg sample in zero applied field using a Quantum Design PPMS.

The dynamic structure factor, $S\text{(}Q,E=\hbar \omega \text{)}$, was measured for the same 5~g $^7$Li-enriched \LMP~sample on the direct geometry time-of-flight IN5 spectrometer at the ILL. Three spectra were collected at 1.8~K with incident energies $E_\textrm{i} = $ 3.55~meV, 5.11~meV, and 14.2 meV. 

Interaction parameters of the spin Hamiltonian 
\begin{equation}
\mathcal{H} = \sum_{\braket{ij}} J_{ij} S_i\cdot S_j
    \label{EQN:1dhamiltonian}
\end{equation}
where the summation is over lattice bonds $\braket{ij}$, were obtained from density-functional theory (DFT) band-structure calculations performed in the FPLO\cite{fplo} and VASP\cite{VASP1,VASP2} codes using the Perdew-Burke-Ernzerhof (PBE)\cite{Perdew1996} flavor of the exchange-correlation potential. Experimental lattice parameters and atomic positions from Ref.\cite{LMP} were used.

The exchange parameters $J_{ij}$ were extracted using two complementary procedures. First, we calculate the hopping integrals between the Mo 4$d$ states using Wannier functions constructed for the uncorrelated (PBE) band structure, and introduce these hoppings into the Kugel-Khomskii model that leads to the magnetic exchange couplings\cite{Mazurenko2006,Tsirlin2011},
\begin{equation}
    J_{ij} = \frac{4[t_{ij}^{(nn)}]^2}{U_\textrm{eff}} - \sum_{m}\frac{4[t_{ij}^{(nm)}]^2J_\textrm{eff}}{(U_\textrm{eff}+\Delta_m)(U_\textrm{eff}-J_\textrm{eff}+\Delta_m)},
\end{equation}
\label{eq:2}
where the first and second terms stand for the antiferromagnetic and ferromagnetic contributions, respectively. Here, $U_\textrm{eff} = 4$~eV is the effective on-site Coulomb repulsion and $J_\textrm{eff} = 0.5$~eV is the effective Hund's coupling in the Mo 4$d$ shell. The hoppings $t_{ij}^{(nn)}$ are between the half-filled states of molybdenum, whereas $t_{ij}^{(nm)}$ involve the higher-lying empty states $m$, and $\Delta_m = \varepsilon_m - \varepsilon_n$ is the crystal-field splitting.

Alternatively, we obtain the exchange couplings by a mapping procedure\cite{Xiang2011,Tsirlin2014} using the total energies of collinear spin configurations evaluated within DFT$+U$. Here, correlation effects in the Mo 4$d$ shell are treated on the mean-field level with the on-site Coulomb repulsion $U_d = 4$~eV, Hund's coupling $J_d = 0.5$~eV, and atomic-limit flavor of the double-counting correction\cite{Abdeldaim2023}.
%%%%%%%%%%%%%%%%%%%%%%%%%%%%%%%%%%%%%%%%%%%%%%%%%%%%

\section{Results}
\label{sec:results}
%%%%%%%%%%%%%%%%%%%%%%%%%%%%%%%%%%%%%%%%%%%%%%%%%%%%
\subsection{Crystal structure}
%%%%%%%%%%%%%%%%%%%%%%%%%%%%%%%%%%%%%
\begin{figure}
\centering
\includegraphics[width = \columnwidth,keepaspectratio]{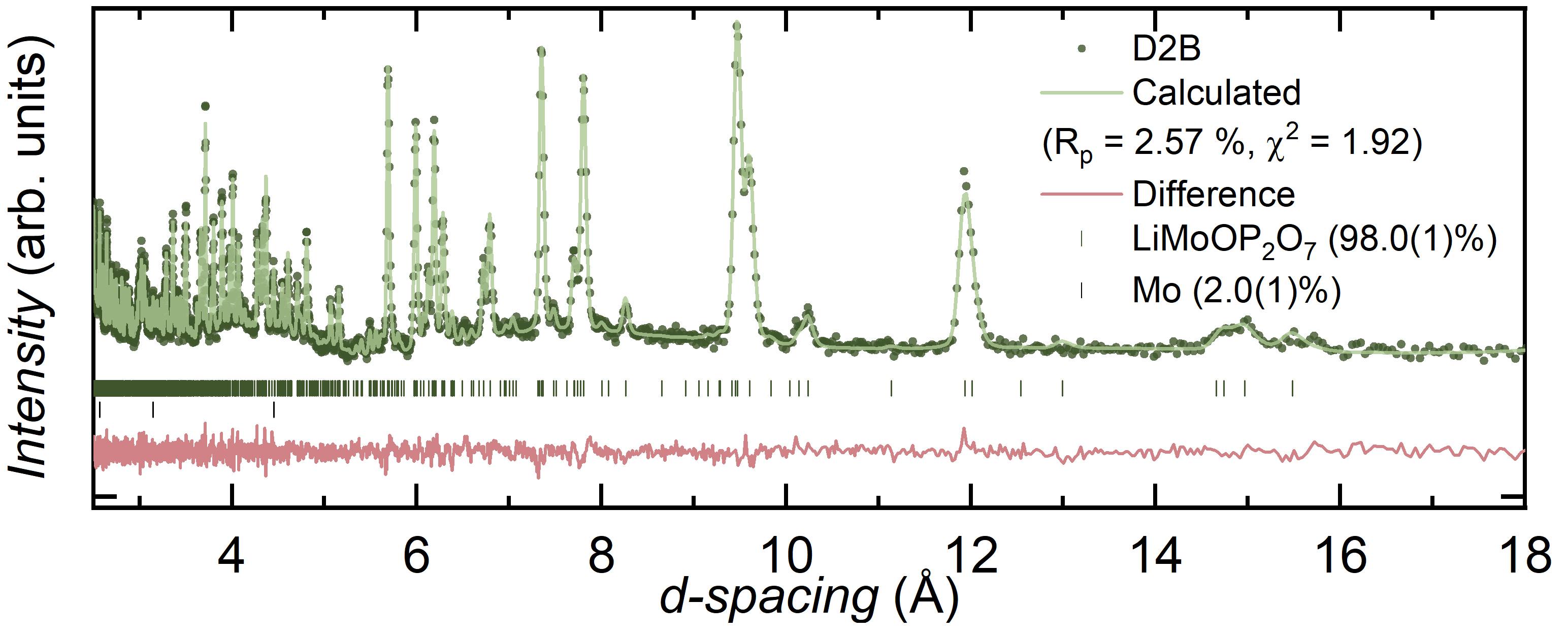}\caption{Rietveld refinement of neutron powder diffraction data collected at 300~K on the D2B diffractometer using the reported $P2_1/n$ structural model\cite{LMP}. The resulting unit cell parameters are $a = 16.015(2)$~\AA, $b = 11.945(1)$~\AA, $c = 9.937(1)$~\AA, and $\beta = 104.688(4)$\degree~with $\chi^2 = 1.92$ and $R_\textrm{p} = 2.57\%$.}
\label{figure2}
\end{figure}
%%%%%%%%%%%%%%%%%%%%%%%%%%%%%%%%%%%%%
\noindent The previously reported $P2_1/n$ structural model of \LMP~\cite{LMP} was confirmed at 300~K by Rietveld refinement of neutron powder diffraction data (Fig.~\ref{figure2}). The lattice parameters and all atomic coordinates were refined. Given the large number of atoms in the unit cell, however, allowing the isotropic thermal displacement parameters to vary rendered the refinement unstable. Hence, the thermal displacement parameters reported from the analysis of single-crystal data measured at 300~K were used as an approximation\cite{LMP}. Bond-length restraints of 1.58~\AA~and 2.1~\AA~were used for all P-O and Li-O bonds, respectively, which were defined based on other P$_2$O$_7$- and LiO$_4$-containing materials\cite{Canadell1997,Li-O}. The resulting model is summarized in Table SI.

In contrast to the quasi-one-dimensional structural model adopted by the rest of the \AMP~series\cite{NMP,Abdeldaim2023,KMP,CMP}, where \textit{A} = Na, K, Cs, \LMP~crystallizes with a different structure containing three unique Mo crystallographic sites\cite{LMP} (Fig. \ref{figure1}(a)). As shown in Fig. \ref{figure1}(b), each Mo site adopts an octahedral coordination that involves four P$_2$O$_7$ groups and one edge-sharing LiO$_4$ tetrahedron. Three of the P$_2$O$_7$ groups share a single corner with the MoO$_6$ octahedron, whereas the fourth is edge-sharing. This coordination leaves one unshared apical oxygen with the characteristically short Mo-O bond of bond length 1.65(1)~\AA~that is observed across the \AMP~family. Empty pores form along these unshared apices and propagate along the crystallographic \textit{a}-axis (Fig. \ref{figure1}(c)). Another tunnel-like structure, occupied by three unique Li cation sites, also forms along the \textit{a}-axis.

%%%%%%%%%%%%%%%%%%%%%%%%%%%%%%%%%%%%%
\begin{figure}
\centering
\includegraphics[width = \columnwidth,keepaspectratio]{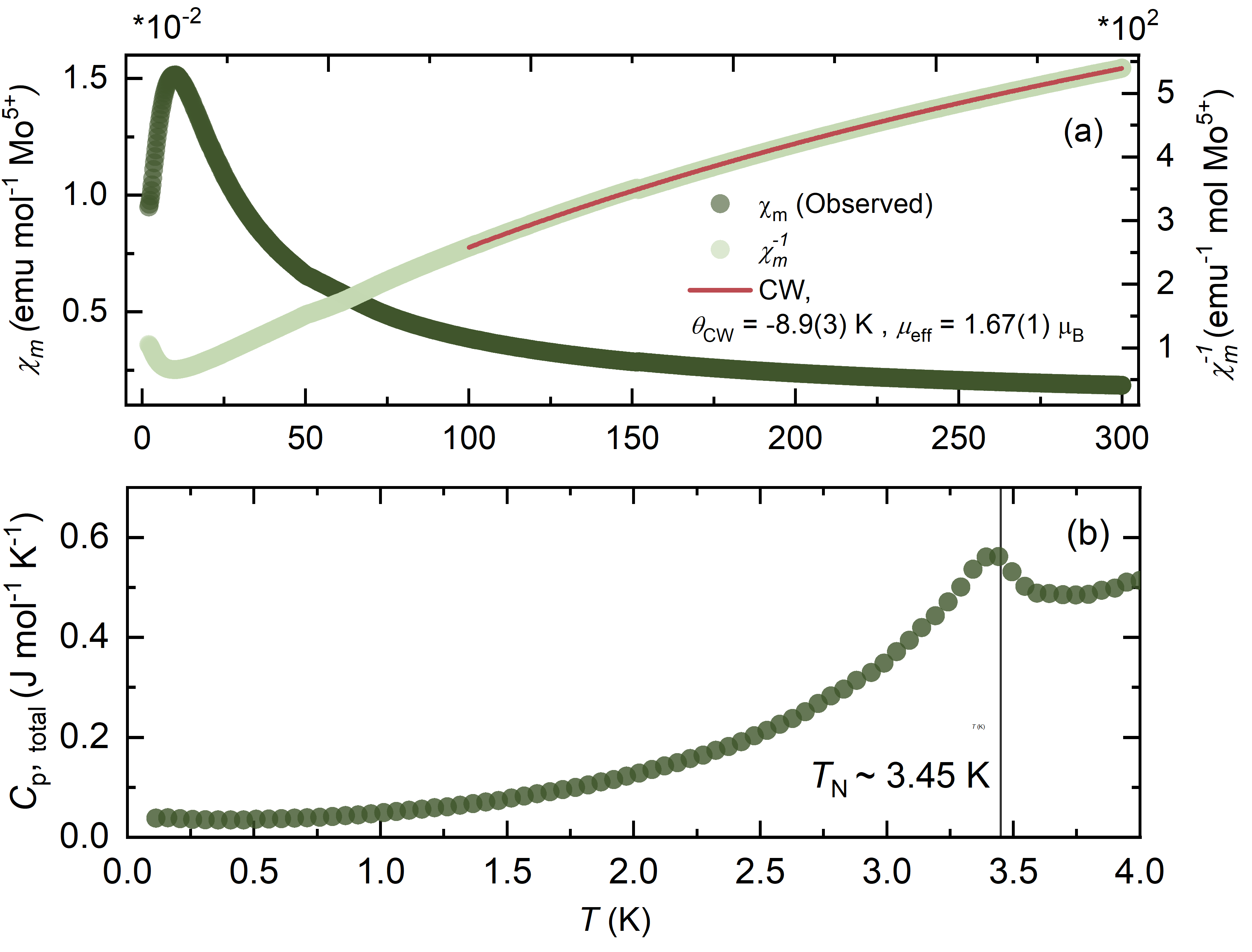}\caption{(a) Temperature-dependent magnetic susceptibility, $\chi_\textrm{m}(T)$, of \LMP~measured in an applied field of 1000~Oe. Above 100~K, $\chi_\textrm{m}^{-1}$(T) is well described by the Curie-Weiss law with model parameters $\theta_\textrm{CW} = -8.9(3)$~K, $C = 0.349(1)$~emu~K~mol$^{-1}$, $g = 1.93$, $\mu_\textrm{eff} = 1.67(1)$ \textmu$_\textrm{B}$, and $\chi_\textrm{0} = 7.64(1) \times~10^{-4}$~emu~mol\textsuperscript{-1}. (b) Temperature dependent zero-field specific heat.}
\label{figure3}
\end{figure}
%%%%%%%%%%%%%%%%%%%%%%%%%%%%%%%%%%%%%

When considering the magnetic sublattice, LMP~could be described as a three-legged spin ladder propagating in the [301] plane (Fig. \ref{figure1}(d-e)). These ladders are bridged by phosphate groups within the \textit{ac} plane, forming a three-dimensional framework (Fig. \ref{figure1}(e)). Finally, consistent with the rest of the alkali metal series, and as confirmed through DFT+\textit{U} calculations (Sec. \ref{sec:DFT}), the shortest Mo-O bond breaks the degeneracy of the $t_{2g}$ manifold into a lower-lying $d_\textrm{xy}$ active magnetic orbital and higher-energy $d_\textrm{xz}$ and $d_\textrm{yz}$ orbitals. The leading magnetic exchange interactions, as such, are most likely those with components in the [301] plane as depicted in Fig. \ref{figure1}(e) and confirmed in Sec. \ref{sec:DFT}.
%%%%%%%%%%%%%%%%%%%%%%%%%%%%%%%%%%%%%%%%%%%%%%%%%%%%
\subsection{Magnetometry and specific heat}
%%%%%%%%%%%%%%%%%%%%%%%%%%%%%%%%%%%%%%%%%%%%%%%%%%%%

\noindent The zero field-cooled temperature-dependent magnetic susceptibility measured in an applied field of 1000 Oe is shown in Fig. \ref{figure3}(a). Above 100~K, $\chi_\textrm{m}(T)$ is well described by the Curie-Weiss law and yields $C = 0.349(1)$~emu~K~mol\textsuperscript{-1} ($g = 1.93$), $\mu_\textrm{eff} = 1.67(1)$~\textmu$_\textrm{B}$, $\theta_\textrm{CW} = -8.9(3)$~K, and $\chi_\textrm{0} = 7.64(1) \times 10^{-4}$~emu~mol\textsuperscript{-1}. The model parameters are suggestive of dominant antiferromagnetic interactions and an effective magnetic moment comparable to the 1.73~\textmu$_\textrm{B}$ expected for a $S = 1/2$ system with a quenched orbital moment, consistent with the electronic structure (Sec. \ref{sec:DFT}). Upon cooling, an anomaly is observed around 45~K. This feature could be assigned to the presence of oxygen in the measurement chamber as it is consistent with the paramagnetic transition for solid O$_2$\cite{kanda1954paramagnetic}. It is also inconsistent with the energy scale of the exchange interactions (Sec. \ref{sec:INS}), and does not correspond to any structural transitions (Sec. \ref{sec:magstr}). It is thus most likely associated with the relatively large temperature-independent background term $\chi_\textrm{0}$. Consistent with $\theta_\textrm{CW} = -8.9(3)$~K, at lower temperatures, a broad feature, centered about 9.5~K, marks the development of short-range correlations. This is followed by an inflection point at $T_\textrm{N} = 3.45~K$ at which the system undergoes long-range magnetic order. As seen in Fig. \ref{figure3}(b), this ordering transition also features in the temperature dependence of the zero-field specific heat. 

\begin{figure}
\centering
\includegraphics[width = \columnwidth,keepaspectratio]{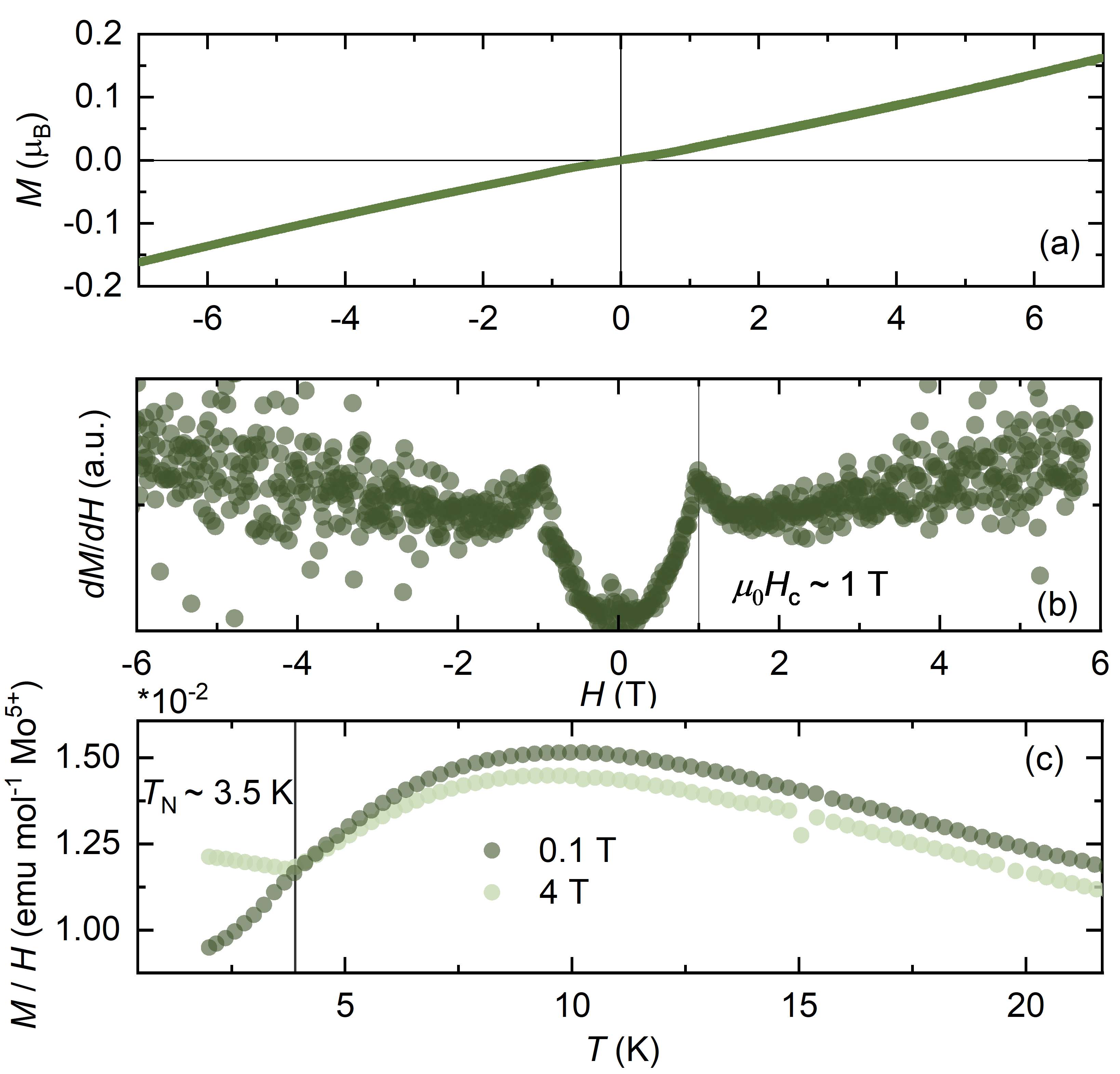}\caption{(a) Magnetization isotherm of \LMP~measured at 1.8 K. (b) Field derivative of the magnetization marking a critical field $H_\textrm{c} = 1$~T. (c) Temperature-dependent $M/H$ curves measured at applied fields of 0.1~T and 4~T.}
\label{figure4}
\end{figure}

The isothermal field-dependent magnetization, $M$(H), is shown in Fig. \ref{figure4}(a). No magnetic hysteresis is observed, suggesting antiferromagnetic ordering with minimal spin-canting, and the moment does not saturate up to an applied field of 7~T. At around 1~T, a field-induced transition is observed as an anomaly in the field derivative of the magnetization (Fig. \ref{figure4}(b)). To examine its effect on the temperature dependence, $\chi_\textrm{m}(T)$ was measured at 4~T. When compared to the curve measured below the critical field, $H_\textrm{c} = 1$~T, $\chi_\textrm{m}(T)$ at 4~T increases at temperatures below $T_\textrm{N}$ (Fig. \ref{figure4}(c)), typical of a spin-flop transition.

%%%%%%%%%%%%%%%%%%%%%%%%%%%%%%%%%%%%%
%%%%%%%%%%%%%%%%%%%%%%%%%%%%%%%%%%%%%

%%%%%%%%%%%%%%%%%%%%%%%%%%%%%%%%%%%%%%%%%%%%%%%%%%%%

\subsection{Magnetic structure}
\label{sec:magstr}

%%%%%%%%%%%%%%%%%%%%%%%%%%%%%%%%%%%%%

\begin{figure}
\centering
\includegraphics[width = \columnwidth,keepaspectratio]{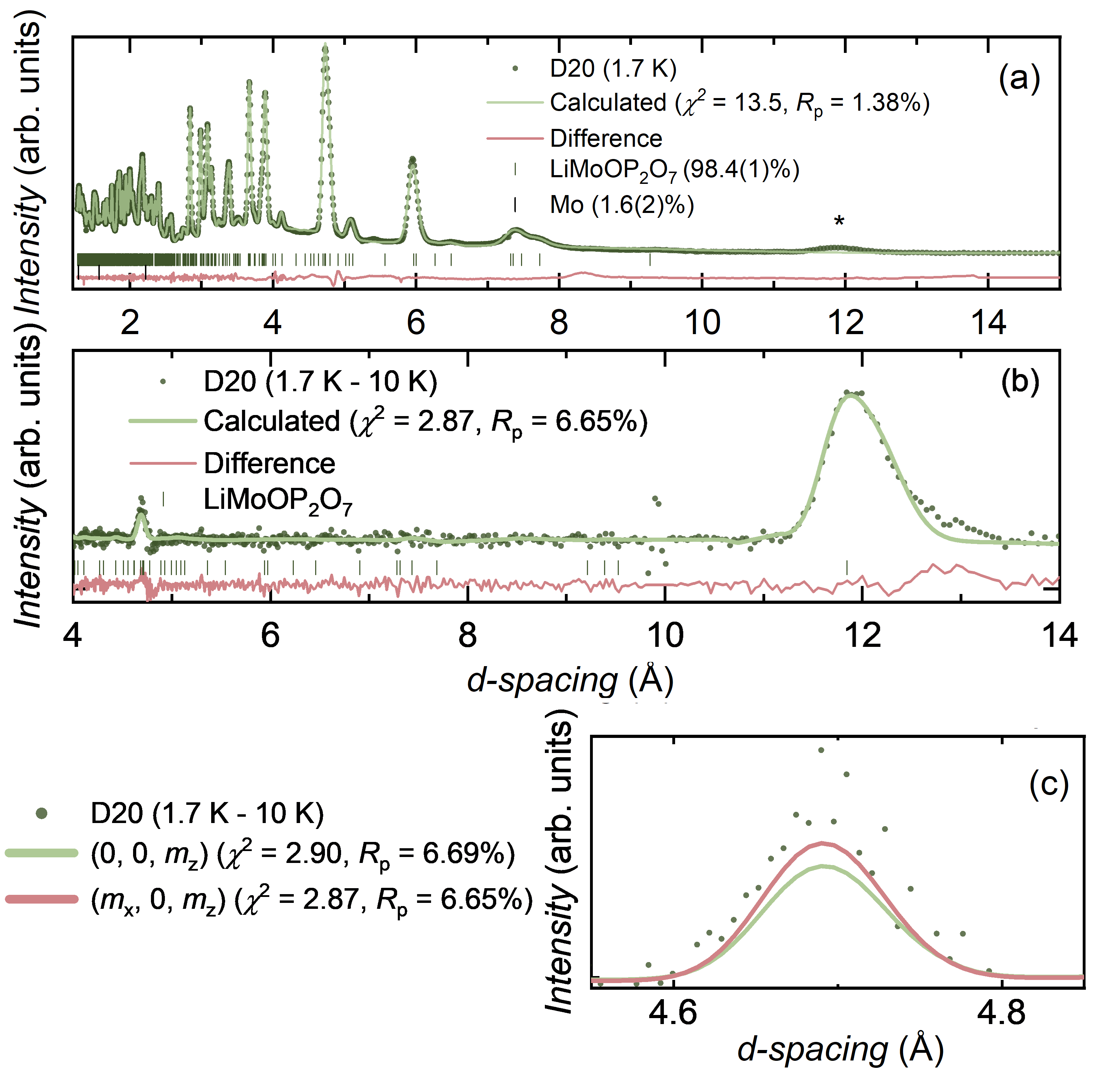}\caption{(a) Rietveld analysis of neutron powder diffraction data measured at 1.7~K using the $P2_1/n$ structural model. The resulting unit cell parameters are $a = 15.978(3)$~\AA, $b = 11.930(1)$~\AA, $c = 9.916(2)$~\AA, and $\beta = 104.686(5)$\degree with $\chi^2 = 13.5$ and $R_\textrm{p} = 1.38\%$. The asterisk highlights the (010) magnetic reflection. (b) Rietveld refinement of magnetic-only scattering at 1.7~K using the $P2_{1}/n^{'}$ magnetic space group model. (c) Comparison of the simulated intensity of the ($\overline{3}11$) magnetic reflection for models in which (green) only $m_z$ or (red) $m_x$ and $m_z$ are refined, respectively.}
\label{figure5}
\end{figure}

\noindent To examine the nature of the magnetic order at 3.5~K, neutron powder diffraction data were collected at 10~K and 1.7~K. No features consistent with a structural transition were observed at 1.7~K, and the $P2_1/n$ structural model was confirmed through Rietveld analysis (Fig. \ref{figure5}(a)). Given the large number of atoms in the unit cell and the poor resolution of the D20 diffractometer at large angles, however, it was not possible to refine all atomic positions and isotropic thermal parameters independently. The isotropic thermal parameters were refined while constrained across like atoms. The resulting model is consistent with an isotropic contraction of the lattice parameters with respect to temperature and the atomic positions and displacement parameters are summarized in Table SII. 

%%%%%%%%%%%%%%%%%%%%%%%%%%%%%%%%%%%%%
%%%%%%%%%%%%%%%%%%%%%%%%%%%%%%%%%%%%%

Subtracting the 10~K data from that collected at 1.7~K yields two magnetic Bragg reflections that are indexed by the commensurate propagation vector, $\kappa = 0$ (Fig. \ref{figure5}(b)). Symmetry analysis using \texttt{BasiReps}\cite{fullprof} and \texttt{MAXMAGN}\cite{maxmagn} results in four irreducible representations, $m\Gamma_{2+}$, $m\Gamma_{2-}$, $m\Gamma_{1+}$ and $m\Gamma_{1-}$ in Miller-Love notation\cite{miller1967tables}, that are compatible with $\kappa$ and the crystal symmetry. Of those, only the magnetic space group $P2_{1}/n^{'}$, in Belov-Neronova-Smirnova notation\cite{belov1957shubnikov}, belonging to the $m\Gamma_{1-}$ representation, is consistent with the experimental reflections. 

As this magnetic space group allows for three Fourier components on each Mo site, a total of 14 models were refined against the data. Each ($m_x$, $m_y$, $m_z$) combination was refined under two conditions. In the first, the Fourier components were constrained to be identical across all three Mo sites such that, for example, $m_x^{\mathrm{Mo1}}=m_x^{\mathrm{Mo2}}=m_x^{\mathrm{Mo3}}$. In the second, all Fourier components were allowed to vary independently on each Mo site (\textit{e.g.}, $m_x^{\mathrm{Mo1}}\neq m_x^{\mathrm{Mo2}}\neq m_x^{\mathrm{Mo3}}$). The observed magnetic reflections could only be captured by models in which the magnetic moments were aligned either along the $c$-direction or within the $ac$ plane, whereas allowing $m_y$ to refine consistently yielded negligible values $(<0.008(80)~\mu_{\mathrm B})$. As seen in Fig. \ref{figure5}(c), among the remaining solutions, allowing for both $m_x$ and $m_z$ to refine results in a slightly better fit ($\chi^2 = 2.87$ and $R_\textrm{p} = 6.65\%$) to the data than allowing for only $m_z$ to refine ($\chi^2 = 2.90$ and $R_\textrm{p} = 6.69\%$). The resulting best fitting model when the magnetic moments are constrained to be equal across all Mo sites yields an ordered magnetic moment that is suppressed from the full value of  $gS = 1$~\textmu$_\textrm{B}$ for $S = 1/2$, with $m_x = 0.21(1)$~\textmu$_\textrm{B}$ and $m_z = 0.67(1)$~\textmu$_\textrm{B}$. We note that equal goodness-of-fit parameters result when $m_x$ and $m_z$ are allowed to vary independently across the three Mo sites. However, given that only two magnetic reflections are seen in the data, these refinements are unstable, and a concrete determination of the moment sizes across the crystallographic sites must await a single-crystal study. 

\begin{figure}
\centering
\includegraphics[width = \columnwidth,keepaspectratio]{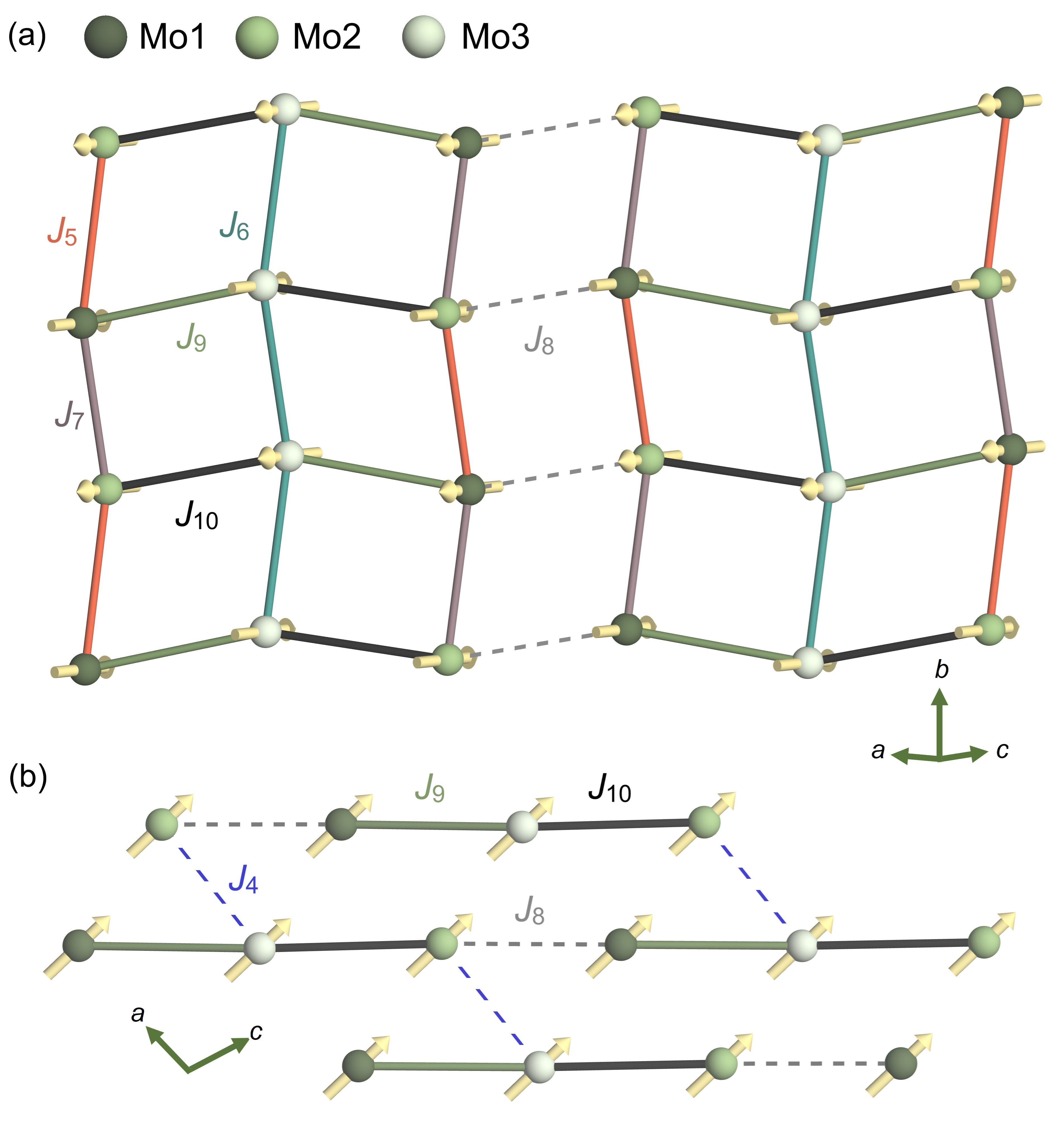}\caption{Magnetic structure and exchange pathways in \LMP~as viewed within the (a) [301] plane and the (b) \textit{ac} plane.}
\label{figure7}
\end{figure}

The resulting magnetic structure is shown in Fig. \ref{figure7}. Here, the moments are aligned antiferromagnetically along the ladder legs and are coupled ferromagnetically across the rungs, where moments predominantly point along the $c$-direction and are canted towards $a$. This antiferromagnetic structure is consistent with the lack of a hysteresis loop in the collected isothermal magnetization curve (Fig. \ref{figure4}(a)). It is further consistent with the resulting model exchange Hamiltonian from both DFT calculations (\ref{sec:DFT}) and analysis of the magnetic excitation spectrum (\ref{sec:INS}) where ladder interactions are predominantly antiferromagnetic whereas all other interactions are ferromagnetic. 

%%%%%%%%%%%%%%%%%%%%%%%%%%%%%%%%%%%%%
%%%%%%%%%%%%%%%%%%%%%%%%%%%%%%%%%%%%%

%%%%%%%%%%%%%%%%%%%%%%%%%%%%%%%%%%%%%%%%%%%%%%%%%%%%

\subsection{Microscopic magnetic model}
\label{sec:DFT}

\begin{figure}
\centering
\includegraphics[width = \columnwidth,keepaspectratio]{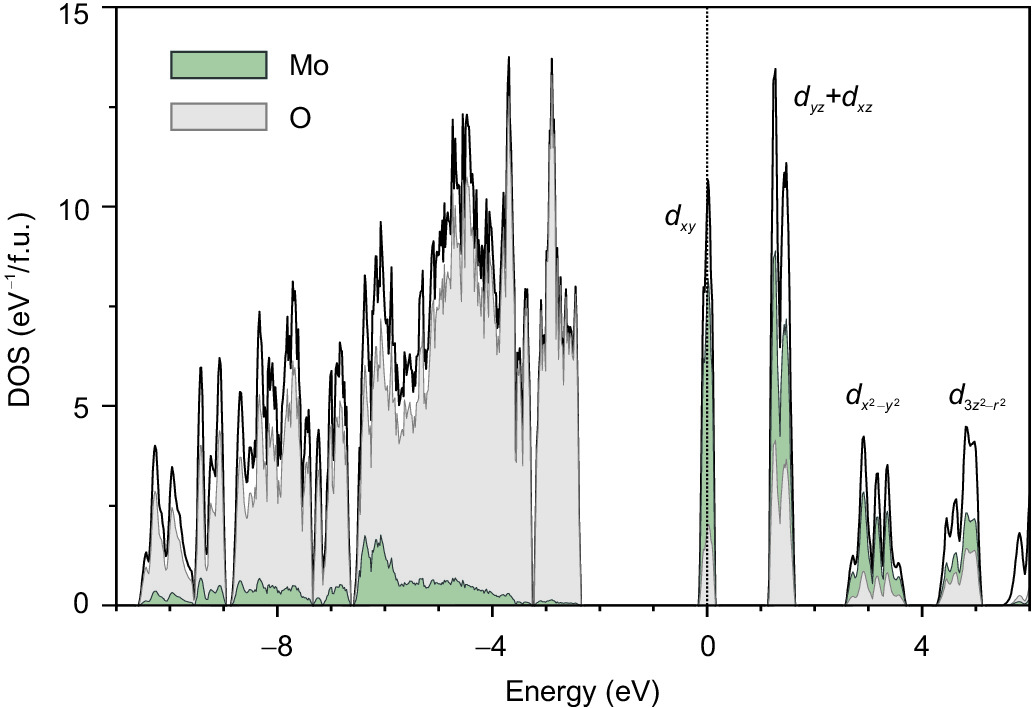}\caption{PBE density of states (DOS) highlighting the crystal-field splitting of the Mo 4\textit{d} states. The Fermi level is at zero energy. Note that the energy spectrum is metallic because strong electronic correlations in the Mo 4$d$ shell are not taken into account on the PBE level.}
\label{figure6}
\end{figure}

\noindent Having determined the magnetic structure of \LMP, we now turn to the electronic structure. The PBE band structure of \LMP~(Fig. \ref{figure6}) shows strong similarities to that of KMoOP$_2$O$_7$\cite{Abdeldaim2023}, despite the two compounds featuring markedly different crystal structures. The Mo 4$d$ bands span the energy range between $-0.2$ and 5.2~eV and reveal a pronounced crystal-field splitting. The lowest-energy orbital is $d_{xy}$ assuming $z$ as the direction of the short Mo-O bond in the MoO$_6$ octahedron. Similar crystal-field splittings are found for all three Mo sites (Table \ref{tab:orben}).

By choosing $d_{xy}$ as the magnetic orbital, we evaluate the ferromagnetic (FM) and antiferromagnetic (AFM) contributions to the exchange interactions using Eq. \ref{eq:2} and concurrently calculate the total exchange couplings from DFT+$U$ (Table \ref{tab:DFTexch}). Both methods arrive at consistent results and reveal three dominant AFM exchange parameters along with several weaker FM couplings. Remarkably, the shortest Mo--Mo distances of 5.0--5.3 \AA~are not responsible for any significant interactions, whereas the dominant interactions $J_5$ to $J_7$ involve Mo atoms that are further apart.

The spin lattice of \LMP~is shown in Fig. \ref{figure7} and can be represented by three-legged spin ladders stretched along the $b$ direction. The couplings $J_5$ to $J_7$ constitute the AFM legs of the ladders, while the couplings $J_9$ and $J_{10}$ are the FM rungs. The weaker FM coupling $J_8$ connects the ladders into layers arranged perpendicular to the [301] plane. Finally, $J_4$ is the dominant interlayer coupling.

%%%%%%%%%%%%%%%%%%%%
\begin{table}[]
\label{LMP_crystalfield}
\centering
\caption{Orbital energies (in eV) extracted from Wannier fits to the PBE band structure of LiMoOP$_2$O$_7$.}
    \renewcommand{\arraystretch}{1.2} % 
    \setlength{\tabcolsep}{9.5pt}
\begin{tabular}{cccccc}
\hline
\hline
     & $xy$ & $yz$ & $xz$ & $x^2-y^2$ & $3z^2-r^2$ \\
 Mo1 & 0.07 & 1.43 & 1.39 & 3.08      & 4.73       \\
 Mo2 & 0.08 & 1.43 & 1.34 & 3.10      & 4.73       \\
 Mo3 & 0.02 & 1.28 & 1.36 & 3.13      & 4.76       \\
\hline \hline
\end{tabular}
\label{tab:orben}
\end{table}
%%%%%%%%%%%%%%%%%%%%%%%%%

%%%%%%%%%%%%%%%%%%%%%%%%%

\begin{table}[]
\centering
\caption{Summary of exchange couplings in LiMoOP$_2$O$_7$, the Mo--Mo distances $d$, AFM and FM exchange couplings derived from the Kugel-Khomskii model (Eq. \ref{eq:2}), total exchange couplings $J_{ij}$ obtained by the DFT+$U$ mapping analysis, and linear spin wave theory fitted model parameters $J_{ij}^{\textrm{LSWT}}$. Note that $J_{ij} \neq J_{ij}^\textrm{AFM} + J_{ij}^\textrm{FM}$ as individual contributions and total exchange couplings are obtained by different methods. All exchange parameters are given in units of meV.}
    \renewcommand{\arraystretch}{1.5} % 
    \setlength{\tabcolsep}{3.2pt}
\begin{tabular}{rrrrrrr}
\hline
\hline
       & $d_{\rm Mo--Mo}$ (\AA) & & $J_{ij}^\textrm{AFM}$  & $J_{ij}^\textrm{FM}$ & $J_{ij}$  & $J_{ij}^{\textrm{LSWT}}$ \\\hline
$J_1$  & 5.000 & Mo1--Mo2  & 0.0  & $-0.03$  & $0.07$ & \\
$J_2$  & 5.160 & Mo2--Mo3  & 0.20  & $-0.13$ & $-0.13$ & \\
$J_3$  & 5.278 & Mo1--Mo3  & 0.03  & $-0.06$ & $-0.13$ & \\ 
$J_4$  & 5.425 & Mo1--Mo2  & 0.03  & $-0.23$ & $-0.26$ & $-0.46$ \\
$J_5$  & 5.896 & Mo2--Mo3  & 1.69  & $-0.05$ & 1.36   & 1.76 \\
$J_6$  & 6.095 & Mo1--Mo1  & 1.86  & $-0.13$ & 1.14   & 1.81 \\
$J_7$  & 6.234 & Mo2--Mo3  & 1.69  & $-0.06$ & 1.53   & 1.72 \\
$J_8$  & 6.238 & Mo2--Mo3  & 0.20  & $-0.42$ & $-0.35$ & $-0.33$ \\
$J_9$  & 6.333 & Mo1--Mo3  & 0.06  & $-0.41$ & $-1.02$ & $-0.34$ \\
$J_{10}$ & 6.339 & Mo1--Mo2  & 0.22  & $-0.35$ & $-0.48$ & $-0.64$ \\
$J_{11}$ & 7.229 & Mo3--Mo3  & 0.26  & $-0.03$ & 0.22 &  \\
$J_{12}$ & 7.332 & Mo1--Mo2  & 0.29  & $-0.04$ & $0.12$ & \\
\hline \hline
\end{tabular}
\label{tab:DFTexch}
\end{table}
%%%%%%%%%%%%%%%%%%%

%%%%%%%%%%%%%%%%%%%%%%%%%%%%%%%%%%%%%
%%%%%%%%%%%%%%%%%%%%%%%%%%%%%%%%%%%%%

%%%%%%%%%%%%%%%%%%%%%%%%%%%%%%%%%%%%%
\begin{figure*}
\centering
\includegraphics[width = 0.9\textwidth,keepaspectratio]{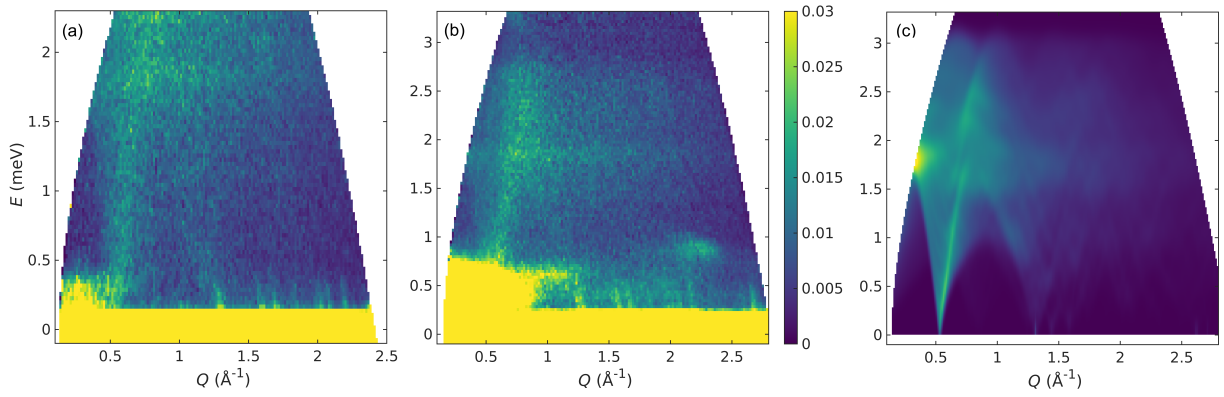}\caption{Experimental dynamic structure factor measured on the IN5 spectrometer at 1.8~K with incident neutron energies (a) $E_\textrm{i} = 3.55~$meV and (b) $E_\textrm{i} = 5.11~$meV. (c) Linear spin wave theory simulated spectrum using the best fitting model parameters $J_4 = -0.46$, $J_5 = 1.76$, $J_6 = 1.81$, $J_7 = 1.72$, $J_8 = -0.33$, $J_9 = -0.34$, and $J_{10} = -0.64$~meV.}
\label{Figure8}
\end{figure*}
%%%%%%%%%%%%%%%%%%%%%%%%%%%%%%%%%%%%%

Table \ref{tab:DFTexch} lists all the exchange parameters that exceed 0.05~meV in magnitude calculated either using Eq. \ref{eq:2} or the mapping analysis. The couplings $J_4$ to $J_{10}$ are non-frustrated, as shown in Fig. \ref{figure7}. The respective minimum magnetic model is sufficient to capture the main features of the magnetic ground state of \LMP. It reproduces the $\kappa = 0$ magnetic order and results in Curie-Weiss temperatures of $\theta_1 = -1.5$~K (Mo1), $\theta_2 = -5.2$~K (Mo2), and $\theta_3 = -4.4$~K (Mo3) for the individual Mo sites. The average Curie-Weiss temperature of $-3.7$~K is on the same energy scale as the experimental value of $-8.9(3)$~K.

\subsection{Inelastic neutron scattering}
\label{sec:INS}
\noindent With the minimal model spin Hamiltonian established, we next determine the exchange parameters experimentally by examining inelastic neutron scattering spectra. The dynamic structure factor, $S\text{(}Q,E=\hbar \omega \text{)}_{\textrm{exp}}$, measured at 1.8~K with $E_\textrm{i} = 3.55$ and $5.11$~meV is shown in Fig. \ref{Figure8} (a) and (b), respectively. As spurious scattering arising from the sample environment is seen at low energies and small angles, we limit our analysis to $E > 1$~meV.
%%%%%%%%%%%%%%%%%%%%%%%%%%%%%%%%%%%%%
\begin{figure}
\centering
\includegraphics[width = \columnwidth,keepaspectratio]{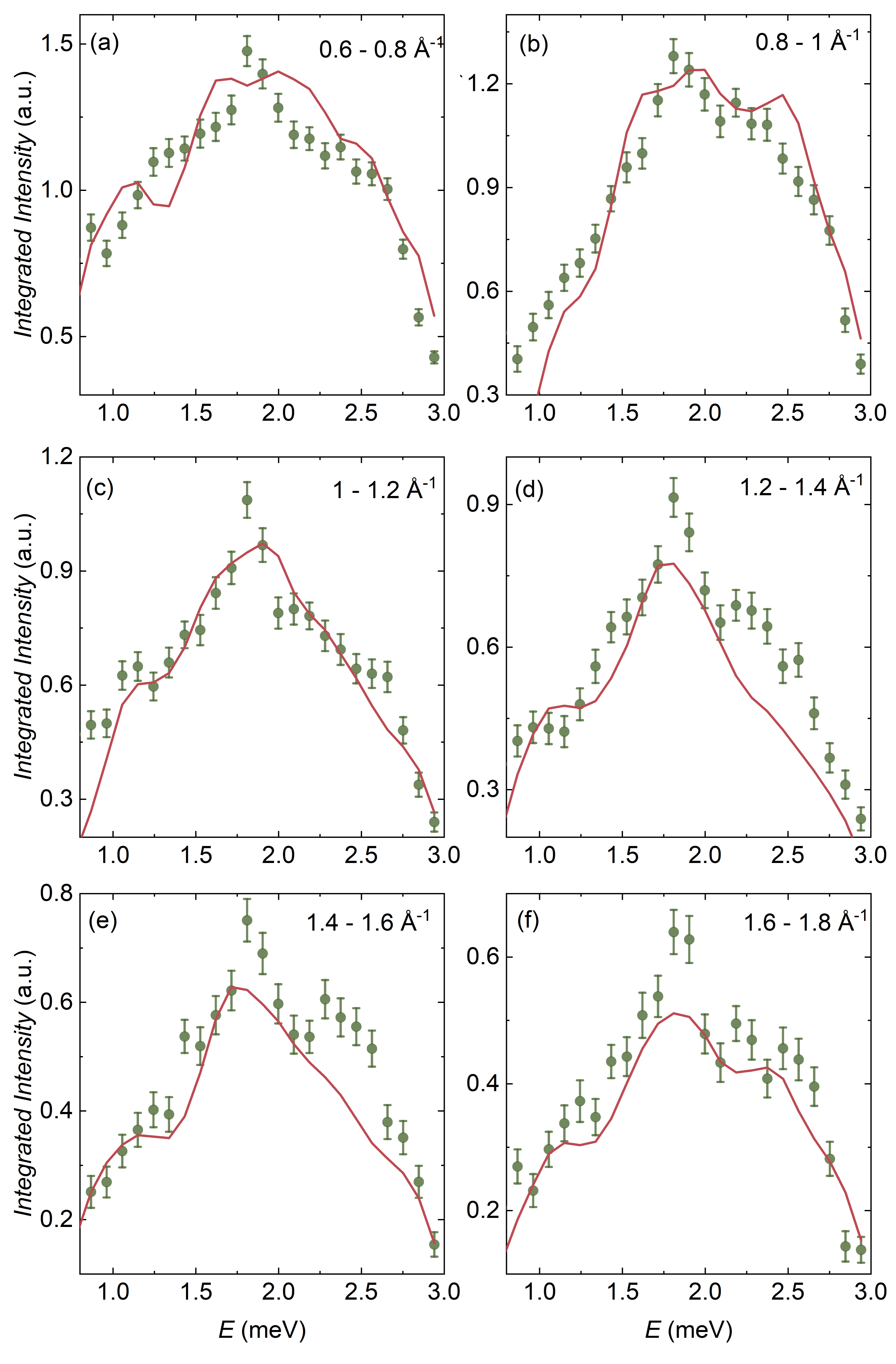}\caption{(a-f) Constant momentum transfer $\lvert Q \rvert$ cuts fit to the linear spin wave theory calculated dynamic structure factor with $J_4 = -0.46$, $J_5 = 1.76$, $J_6 = 1.81$, $J_7 = 1.72$, $J_8 = -0.33$, $J_9 = -0.34$, and $J_{10} = -0.64$~meV.}
\label{Figure9}
\end{figure}
%%%%%%%%%%%%%%%%%%%%%%%%%%%%%%%%%%%%%

The experimental data were fit with the seven leading exchange parameters, $J_4-J_{10}$, using linear spin-wave theory (LSWT) as implemented in \texttt{SpinW}\cite{SpinW}. All the calculated dynamic structure factor intensities were powder averaged and convoluted with the instrumental resolution. Six constant momentum transfer $\lvert Q\rvert$ cuts, integrated over $\delta Q = 0.1$~\AA\textsuperscript{-1}, and centered about 0.7, 0.9, 1.1, 1.2, 1.4, and 1.6~\AA\textsuperscript{-1} were used (Fig. \ref{Figure9}). For each cut, a constant background term was subtracted and a single overall amplitude was fitted. To optimize the model, a two-stage procedure was used. First, a 2 meV bounded parameter space, centered about the DFT+$U$ calculated exchange parameters, was explored using a surrogate optimization algorithm with 5000 evaluations. This was repeated 20 times to ensure that the parameter space minimum was consistent across all optimizations, and its results are reported in Table SIII. The best fitting solutions obtained from this search were subsequently refined using a local pattern-search algorithm, with the variation of each parameter constrained to $\pm 15\%$ of its initial value. Figure \ref{Figure8}(c) shows a representative simulated $S\text{(}Q,E=\hbar \omega \text{)}_{\textrm{LSWT}}$ for one of the best fitting solutions, presented in Table \ref{tab:DFTexch}, with fits to the constant momentum transfer $\lvert Q\rvert$ cuts shown in Fig. \ref{Figure9}.  

On examining the resulting model parameters (Table SIII), we find that the values of $J_4-J_5$ and $J_8-J_{10}$ remain consistent across the different optimization runs and are well constrained by the measured spectrum, with $-0.59<J_4<-0.43$, $1.51 <J_6 <1.90$, $-0.38<J_8<-0.29$, $-0.39< J_9 < -0.29$, and $ -0.65 <J_{10} < -0.55$, whereas $1.25<J_5<1.84$ and $1.64<J_7<2.35$ vary more considerably. To establish the origin of this variation, we have performed a grid search where $J_5-J_7$ were varied in $20 \times 20 \times 20$ linearly spaced steps between $0.90<J_5<2.70$, $0.78<J_6<2.58$, and $1.26<J_7<3.05$ while fixing $J_4 = -0.53$, $J_8 = -0.34$, $J_9 = -0.32$, and $J_{10} = -0.60$, to one of the optimized model parameters. The resulting fit residual landscape is shown in Fig. S1 where both a low residual region and a clear negative correlation between $J_5$ and $J_7$ are seen. As such, $J_5$ and $J_7$ cannot be constrained with the current dataset, and further insight would require, for example, a single-crystal inelastic neutron scattering measurement. Regardless, the resulting model parameters well describe the measured spectrum and are consistent with the DFT results; the largest exchange parameters are AFM and correspond to couplings along the ladder legs ($J_5-J_7$), while interactions along the rungs are weaker and FM ($J_8-J_{10}$), with a significant FM interlayer coupling ($J_4$). 

\section{Discussion}
\label{sec:discussion}

\noindent Despite the structural complexity of \LMP~and the powder averaging of the experimental data, our analysis of the experimental INS spectra has allowed for a seven-parameter exchange Hamiltonian to be considerably constrained within the model's parameter space. The remaining ambiguity lies in the relative ratio of $J_5$ to $J_7$ which form an alternating spin chain of Mo2 and Mo3 that share a similar exchange pathway and are strongly anti-correlated. 

Some additional qualitative considerations may nevertheless further constrain the parameter space. For an isolated $S = 1/2$ AFM spin chain with alternating exchange interactions, the spin-spin interactions form coupled dimers, and the excitation spectrum is gapped to the first excited state\cite{Cross1979}. Taking $\alpha = J_7/J_5 = 0.9$ and $J_7 \approx 1.8$~meV, such a gap would be as large as $\Delta \approx 0.35$~meV\cite{Tennant1999,Papenbrock2003}, and one could naively expect a vestige of it to remain even in a more complex, three-dimensional network. We find no such gap within the experimental resolution $\delta E = 0.13$~meV of the 3.55 meV experimental INS spectrum. Additionally, in solutions with $\alpha \approx 1$, $J_5$ and $J_7$ converge to a similar energy scale to $J_6$. Solutions with a strong $J_5$ and $J_7$ alternation would require similar exchange pathways to result in markedly different exchange parameters with no evident structural or orbital origins. Overall, these considerations favor the region of the parameter space where $J_5 \approx J_6 \approx J_7 \approx 1.8$~meV. 

We now turn to the modeled spin Hamiltonian and its magnetic ground state. In an ideal Heisenberg $S = 1/2$ three-legged spin-ladder with dominant antiferromagnetic leg and weaker ferromagnetic rung interactions, the excitation spectrum is gapless and the system does not undergo long-range magnetic order\cite{Oitmaa1996}. In \LMP, we find a significant interlayer interaction $\lvert J_4/J_6 \rvert \approx 0.25$ that drives the system away from the ideal model. Additionally, both Dzyaloshinskii–Moriya interactions and off-diagonal terms are symmetry allowed in the exchange tensor. Given how well the model captures the experimental features in $S\text{(}Q,E=\hbar \omega \text{)}_{\textrm{exp}}$, however, these are most likely small, and further modeling would require single-crystal INS data.

Both the DFT$+U$ calculations and the fits to our inelastic neutron scattering data support a model spin Hamiltonian where the relative strengths of the exchange interactions could not be easily inferred from interatomic distances alone. This is similarly found in VODPO$_4\cdot \frac{1}{2}$ D$_2$O\cite{Tennant1997}, KTi(SO$_4$)$_2$\cite{Nilsen2015}, and BaMo(PO$_4$)$_2$\cite{Abdeldaim2019}, for example, and underlines the role of the magnetic orbital in selecting the low-dimensional subunit from the underlying crystal structure. Indeed, in V$^{4+}$ phosphates, the strength of the magnetic interactions is usually determined by the positions of the half-filled $d_{xy}$ orbitals and by the suitable O-O links through edges of the PO$_4$ tetrahedra\cite{Roca1998}. 

A similar reasoning can be applied to \LMP. This can be seen in Fig. \ref{figure1}(e) where the MoO$_6$ octahedra can be viewed as square pyramids, with the magnetic $d_{xy}$ orbital lying in the basal plane. One sees that the intralayer couplings $J_5$ to $J_{10}$ exploit all possible pathways provided by the PO$_4$ tetrahedra. The interlayer interactions $J_3$ and $J_4$ similarly go through PO$_4$ tetrahedra, whereas any further interlayer couplings that could arise from the shortest Mo--Mo distances of 5.0$-$5.3~\AA~do not involve any pathways through the PO$_4$ tetrahedra. This absence of suitable links renders the corresponding superexchange pathways inefficient and makes the respective couplings weak. The splitting of the complex three-dimensional crystal structure of \LMP~into low-dimensional magnetic units is thus prescribed by the positions of the MoO$_5$ square pyramids and their links via PO$_4$. 

\section{Conclusions}
\noindent In summary, we have presented a comprehensive experimental investigation of the $S = 1/2$ Mo\textsuperscript{5+} system, \LMP. By refining neutron powder diffraction data, we confirm the $P2_1/n$ structural model of the system, and find no structural transitions down to 1.8~K. Magnetic susceptibility, isothermal magnetization, and specific heat measurements suggest an antiferromagnetic long-range ordering transition at $T_\textrm{N} \approx 3.5$~K and a field-induced transition at $H_\textrm{c} = 1$~T. Analysis of NPD data confirms this ordering temperature, and the magnetic structure is identified as collinear with magnetic moments aligned within the $ac$ plane. DFT$+U$ calculations suggest similar crystal-field splitting across all Mo crystallographic sites and $d_\textrm{xy}$ as an active magnetic orbital. The role of this active magnetic orbital in yielding larger exchange interactions across longer exchange pathways is discussed, and the minimal model spin Hamiltonian is established. By using linear spin-wave theory to analyze the powder magnetic excitation spectrum measured by inelastic neutron scattering, we determine the seven leading exchange parameters in the spin Hamiltonian. The resulting model suggests that \LMP~can be described as a three-legged $S = 1/2$ ladder with dominant antiferromagnetic ladder interactions, ferromagnetic rung exchanges, and a ferromagnetic interlayer coupling. 
\label{sec:conclusions}

\section{Acknowledgements}
\noindent Provision of a Ph.D. studentship to A. H. A. by the University of Birmingham and the Science and Technology Facilities Council (STFC) is gratefully acknowledged. The authors are also grateful to Dr. Gavin Stenning for aiding with magnetic susceptibility and specific heat measurements at the Materials Characterization Laboratory at ISIS. D. Badrtdinov (Radboud U.) is gratefully acknowledged for calculating the Mo\textsuperscript{5+} covalent form factor. The authors acknowledge the Institut Laue-Langevin (ILL) for beamtime allocation under the experiment codes 5-31-2822\cite{D20} and EASY-844\cite{IN5} for the D20, D2B, and IN5 experiments.
\bibliography{References}{}

\end{document}